\documentclass[9pt,conference]{IEEEtran}
\IEEEoverridecommandlockouts
\usepackage{cite}
\usepackage{amsmath,amssymb,amsfonts}
\usepackage{algorithmic}
\usepackage{graphicx}
\usepackage{textcomp}
\usepackage{xcolor}
\usepackage{subcaption}

\usepackage{hyperref}

\usepackage{color,xcolor}
\usepackage{epsfig}
\usepackage{graphicx}

\usepackage{adjustbox}
\usepackage{array}
\usepackage{booktabs}
\usepackage{multirow}

\usepackage{amsmath,amsfonts,amsthm,amssymb}
\usepackage{bm}
\usepackage{nicefrac}
\usepackage{microtype}

\usepackage{changepage}
\usepackage{extramarks}
\usepackage{fancyhdr}
\usepackage{lastpage}
\usepackage{setspace}
\usepackage{soul}
\usepackage{xspace}
\usepackage{multirow}
\usepackage{multicol}

\def\BibTeX{{\rm B\kern-.05em{\sc i\kern-.025em b}\kern-.08em
    T\kern-.1667em\lower.7ex\hbox{E}\kern-.125emX}}
    
\IEEEoverridecommandlockouts
\IEEEpubid{\makebox[\columnwidth]{ 978-1-6654-3274-0/21/\$31.00~\copyright2021~IEEE \hfill}
\hspace{\columnsep}\makebox[\columnwidth]{ }}

\begin{document}

\title{Invited:Hardware-aware Real-time Myocardial Segmentation Quality Control in Contrast Echocardiography}

\author{
Dewen Zeng$^{1}$ \quad
Yukun Ding$^{1}$ \quad
Haiyun Yuan$^{2}$ \quad
Meiping Huang$^{3}$ \quad
Xiaowei Xu$^{3}$ \quad  \\
Jian Zhuang$^{3}$ \quad
Jingtong Hu$^{4}$ \quad
Yiyu Shi$^{1}$ \\
$^{1}$ Department of Computer Science and Engineering, University of Notre Dame, Notre Dame, IN, USA\\
$^{2}$ Department of Cardiology, Boston Children’s Hospital, Harvard Medical School, Boston, MA, USA\\
$^{3}$ Guangdong Cardiovascular Institute, Guangdong Provincial People's Hospital, Guangzhou, China \\
$^{4}$ Department of Electrical and Computer Engineering, University of Pittsburgh, Pittsburgh, PA, USA \\
}

\maketitle

\begin{abstract}

Automatic myocardial segmentation of contrast echocardiography has shown great potential in the quantification of myocardial perfusion parameters.
Segmentation quality control is an important step to ensure the accuracy of segmentation results for quality research as well as its clinical application.
Usually, the segmentation quality control happens after the data acquisition. At the data acquisition time, the operator could not know the quality of the segmentation results. On-the-fly segmentation quality control could help the operator to adjust the ultrasound probe or retake data if the quality is unsatisfied, which can greatly reduce the effort of time-consuming manual correction. However, it is infeasible to deploy state-of-the-art DNN-based models because the segmentation module and quality control module must fit in the limited hardware resource on the ultrasound machine while satisfying strict latency constraints. In this paper, we propose a hardware-aware neural architecture search framework for automatic myocardial segmentation and quality control of contrast echocardiography. 
We explicitly incorporate the hardware latency as a regularization term into the loss function during training.
The proposed method searches the best neural network architecture for the segmentation module and quality prediction module with strict latency. 

\end{abstract}


\section{Introduction}
Deep learning models have shown remarkable success in automatic myocardial and left ventricle segmentation for contrast echocardiography \cite{li2017fully,li2021deep, dong2018voxelatlasgan}.
After the images are segmented by deep neural networks (DNNs), the segmentation results can be used for further medical analysis such as quantifying myocardial perfusion parameters \cite{li2021deep} and left ventricle ejection fraction \cite{li2017fully}.
However, even the state-of-the-art DNNs will occasionally fail due to low image quality or unexpected features of new data.
In order to ensure accurate medical analysis, segmentation quality control needs to be employed to assess the segmentation quality and detect failure cases.
Traditionally, both the segmentation module and the quality control module are applied after the data acquisition 
\cite{li2017fully, roy2018inherent, robinson2019automated}.
Even though such a framework can filter out good images for further analysis.
One underlying problem is that once all the acquired images of a patient are low quality, the whole framework would not work.
Therefore, it is desirable to design an on-the-fly segmentation and quality control framework to help the operator adjust the ultrasound probe or retake data if the image quality is unsatisfied during data acquisition.

However, building such a framework is non-trivial.
It requires deploying the segmentation model and quality control model on the ultrasound machine with limited hardware resources, while satisfying the real-time requirement.
Existing segmentation works mostly focus on handcrafted neural network architectures to improve the segmentation accuracy \cite{li2021deep, chen2017deeplab}.
In the meantime, some segmentation quality control works try to add additional cues such as shape and appearance feature \cite{ruijsink2020fully} 
or uncertainty \cite{hoebel2020exploration, jungo2018uncertainty} 
to guide quality prediction.
All of these works are computationally expensive that cannot achieve 
the latency requirement ($<$50ms) to avoid noticeable visual lag \cite{annett2014low}.
To enable real-time image segmentation and quality control, there are some works trying to solve one of them.
For example, \cite{wang2019msu} proposed to model the 3D MRI videos as multiscale
canonical form distributions and use a parallel statistical U-net to efficiently process these distributions and boost the inference speed.
\cite{robinson2018real} used a light-weight CNN to predict the Dice of cardiovascular MR segmentation.
The problem is that these methods are designed either from the segmentation perspective or from the quality control perspective, which may not the optimal solution when combined together because 
the performance of these two modules are highly correlated.
In this work, we argue that by jointly optimizing these two modules, we can achieve both high accuracy and low latency for the entire framework.

Neural architecture search (NAS) is a promising technique to achieve this.
Recently, many works have empirically shown that NAS can jointly identify the best architecture and hardware design to maximize both accuracy and hardware efficiency \cite{jiang2019accuracy,wu2019fbnet,cai2018proxylessnas}.
To achieve hardware awareness, the overall latency is estimated on the target platform and jointly optimized with the network accuracy using reinforcement learning (RL) \cite{jiang2019accuracy} or differentiable NAS \cite{wu2019fbnet}.
In this work, we show that by applying the modern hardware-aware NAS technique to the problem of myocardial segmentation and quality control in contrast echocardiography, we can identify the best neural architecture for both segmentation module and quality control module with optimal accuracy and latency performance.
Specifically, we design a supernet that includes a U-net like segmentation network and a Resnet like quality control network.
The segmentation model outputs a segmentation mask and an uncertainty map, which will then be fit into the quality control model to predict the Dice.
We use differentiable NAS \cite{wu2019fbnet} to simultaneously optimize the segmentation loss and quality control loss.
In order to estimate the overall latency of the framework, we profile the latency of each operator used in the search space into a lookup table and model the network latency as a continuous function of the architecture parameters as in \cite{wu2019fbnet}.

The main contributions of this paper are as follows:
\begin{itemize}
    \item We build a neural architecture search framework that can identify the optimal segmentation network and quality control network for contrast echocardiography simultaneously.
    \item We explicitly incorporate the network latency on CPU as a regularization term into the loss function and jointly optimize the segmentation accuracy, quality prediction accuracy and hardware latency during searching.
    \item Experiment results on our collected  Myocardial Contrast Echocardiography (MCE) dataset show that comparing to the handcrafted networks, the searched network can achieve 40.2\% latency reduction while maintaining comparable segmentation and quality prediction accuracy, suggesting that it can be used for real-time myocardial segmentation quality control of contrast echocardiography in clinical routine.
\end{itemize}

\section{Related Work}

\textbf{Segmentation Quality Control.}
Segmentation quality control is a technique to predict the segmentation quality of a given segmentation model (not just the original image) and to detect failure cases that are not suitable for further analysis.
While some works predict segmentation quality using mostly shape and appearance features~\cite{ruijsink2020fully}, DNNs have become popular in segmentation quality prediction along with their adoption for segmentation~\cite{robinson2018real,zhang2019fine}.
Different uncertainty sources and different ways to aggregate pixel/voxel-level uncertainty to segment-level uncertainty are also explored~\cite{jungo2018uncertainty,hoebel2020exploration}. Other notable methods include Monte Carlo dropout~\cite{roy2019bayesian} and Reverse Classification Accuracy (RCA)~\cite{valindria2017reverse} but they are not suitable for real-time applications due to the high computational cost.

\textbf{Neural Architecture Search.} neural architecture search (NAS) is a technique that automatically searches for optimal neural architectures to replace the human effort in designing handcrafted neural architectures.
In existing works, there are mainly three directions in searching an architecture: search space, search strategy, and performance estimation strategy \cite{elsken2019neural}.
The search space defines which architectures can be discovered by the algorithm. 
The search strategy mainly includes reinforcement learning \cite{baker2016designing,jiang2019accuracy}, evolutionary algorithms \cite{real2017large} and gradient-based methods \cite{liu2018darts,cai2018proxylessnas}.
Performance estimation refers to the process of estimating the predictive performance of a candidate architecture on the target dataset.
Recently, gradient-based differentiable NAS has been prevailing because of its high training speed and less hardware requirement \cite{liu2018darts}.
Differentiable NAS models the searching problem as a single training process of an over-parameterized supernet that consists of multiple candidate paths.
Each path is associated with some real-valued architecture parameters that are differentiable to the loss function.
Therefore, standard gradient descent can be used to jointly train the weight parameters and architecture parameters.
Recently, NAS-Unet \cite{weng2019unet}, which uses a cell-based differentiable NAS strategy, shows great success in searching the best neural architecture for medical image segmentation.
However, hardware latency is not considered in their method, which hinders its utilization in real-time situations.

\section{Method}

In this section, we will first introduce the framework overview in Section \ref{section_framework_overview}.
Then, we will discuss the search space of the segmentation network and quality control network in Section \ref{section_segmentation} and Section \ref{section_quality}.
Finally, The hardware-aware loss function will be discussed in Section \ref{section_hardware}.

\subsection{Framework Overview}\label{section_framework_overview}

\begin{figure}[!htb]
	\centering
	\vspace{-12pt}
	\includegraphics[width=0.8\linewidth]{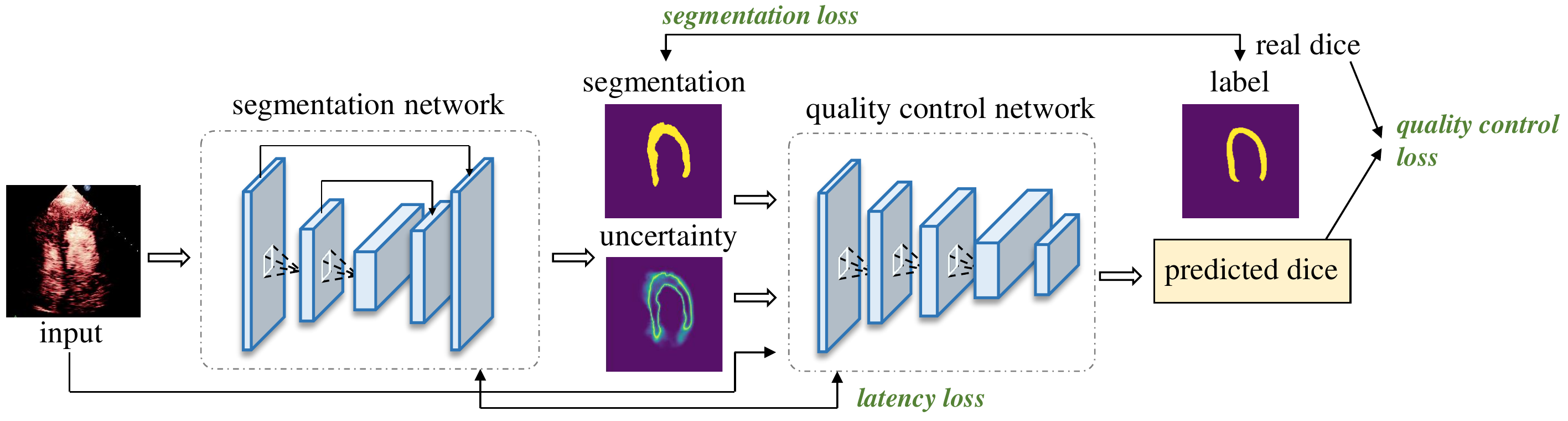}
	\caption{Overview of the proposed NAS framework. The framework jointly searches for the optimal segmentation network and quality control network. In the meantime, the overall hardware latency of the framework is estimated and used as an additional term to the final loss to penalize the network's latency on the target device (CPU).}
	\label{fig:framework_overview}
\end{figure}

Fig.~\ref{fig:framework_overview}  shows the overview of our proposed NAS framework.
The framework consists of a U-net like segmentation network as well as a Resnet like quality control network that needs to be searched.
The segmentation network takes RGB contrast echocardiography as the input and outputs the segmentation result and an uncertainty map.
Then the segmentation result and the uncertainty map, along with the original image are concatenated and fit into the quality control model to predict the estimated Dice.
The ground truth of Dice is computed by using the segmentation result and ground truth mask of myocardial.
The segmentation loss (i.e. cross-entropy loss),  quality control loss (i.e. mean square loss), and the estimated latency of the whole framework will be used in the final loss function during searching.
As such, the segmentation network and quality control network can be optimized and searched simultaneously with better accuracy and latency.

\subsection{Segmentation Network Search Space}\label{section_segmentation}

\begin{figure}[!htb]
	\centering
	\vspace{-12pt}
	\includegraphics[width=0.85\linewidth]{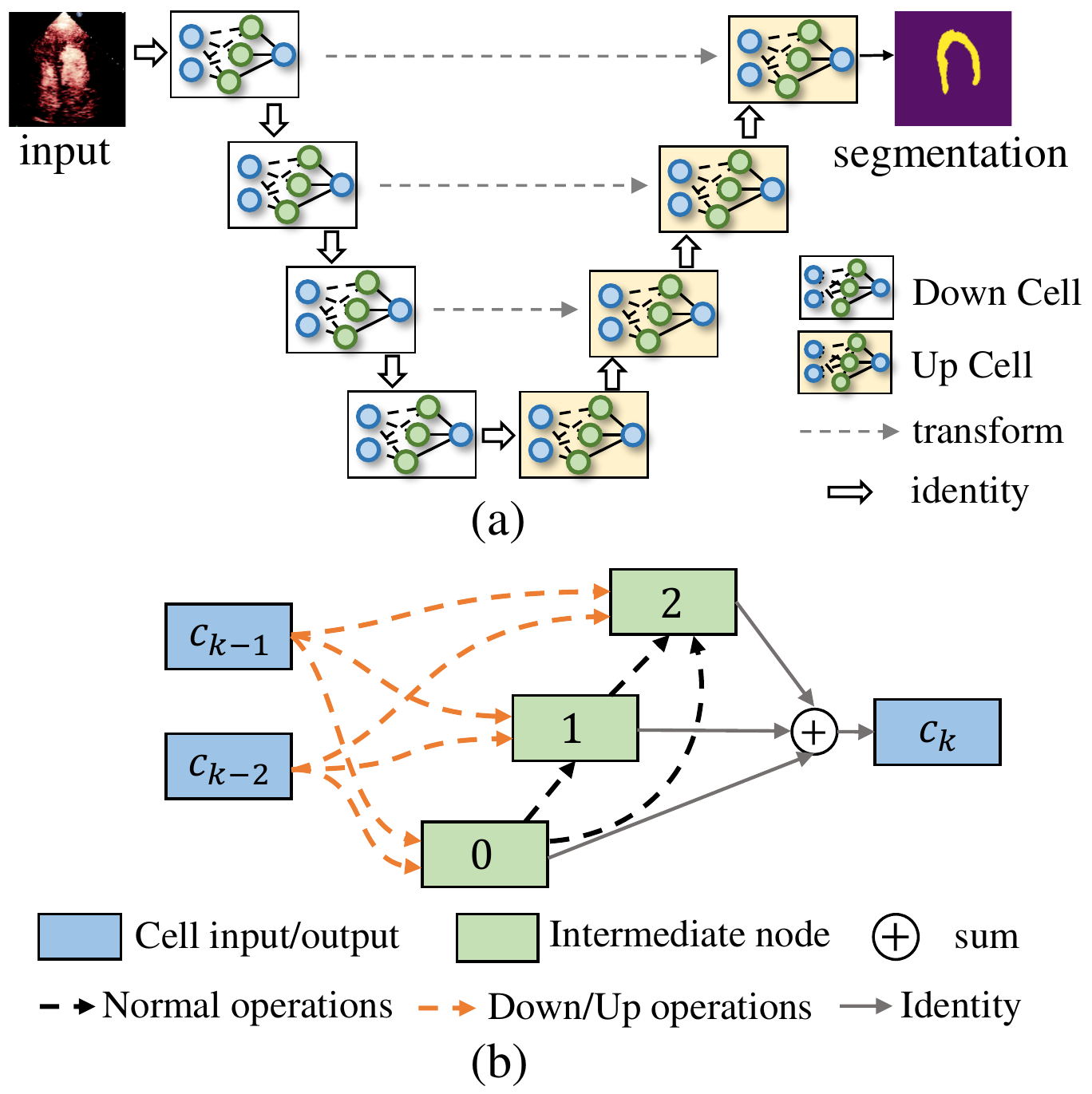}
	\caption{(a) The U-net like segmentation supernet. The network stacks two kinds of cell structures: Down cell and Up cell, which need to be searched. The gray arrow represents the transform operation that belongs the Up cell. (b) Detail cell structure. Each dash arrow represents the operation that needs to be searched from a predefined primitive operation pool. Down and Up operations are used in Down Cell and Up cell respectively.}
	\label{fig:segmentation_network}
\end{figure}

To search for the optimal segmentation network, we use the search space like NAS-Unet \cite{weng2019unet}, which has been proved to be useful for several medical image datasets.
The overall segmentation supernet is shown in Fig.~\ref{fig:segmentation_network}(a).
Two kinds of cell structures called Down cell and Up cell are defined and stacked to form a U-net like backbone.
They share the same network structure as shown in Fig.~\ref{fig:segmentation_network}(b).
The inputs of the cell are defined as the cell out of the previous two cells \cite{liu2018darts}.
The dash edges/lines represent the operations (e.g., convolution operation and pooling operation) that need to be searched between two feature maps (either cell input or intermediate node).
Suppose there are $m$ intermediate nodes inside a cell, the total number of operations need to search is $2m+m(m+1)/2$.
The output of a cell is defined as the sum of all intermediate nodes.

\begin{table}[h]
\centering
\caption{Three types of primitive operation set. For all convolution operations and pooling operations, we use kernel size 3$\times$3.
}
\resizebox{\columnwidth}{!}{
\begin{tabular}{lll}
\hline
Down operations & Up operation & Normal operation \\\hline
down conv  & up conv & conv \\
down dilated conv & up dilated conv & dilated conv \\
down separable conv & up separable conv & seperable conv  \\
max pooling & - & identity \\ 
average pooling & - & zero \\ 
\hline
\end{tabular}}
\label{table:primitive_operations}
\end{table}

Table \ref{table:primitive_operations} shows the three types of primitive operation sets used for searching cells. Following the work of \cite{liu2018darts}, we use 3$\times$3 normal convolution, dilated convolution, and separable convolution.
The down convolution is the stride 2 version of normal convolution, and the up convolution represents the 2d transpose convolution. Pooling operations are only used in down operation set.
Identity and zero operations are only used in normal operation set.

A cell can be viewed as a directed acyclic graph (DAG) with $N$ nodes (including the input and output). 
Since each intermediate node $x_i$ represents a feature map and each edge $(i,j)$ is associated with a primitive operation $op^{(i,j)}$, each intermediate node $x_j$ inside a cell can be computed by all of its previous nodes:
\begin{equation}
    x_j = \sum_{i<j}\sum_{k\in K} p_{\alpha_k^{(i,j)}}\cdot op_k^{(i,j)}(x_i), j>1,
\end{equation}
\begin{equation}
    p_{\alpha_k^{(i,j)}} = \frac{exp(\alpha_k^{(i,j)})}{\sum_{k\in K} exp(\alpha_k^{(i,j)})}.
\end{equation}
where $op_k^{(i,j)}$ is the $k$-th primitive operation from $K$ candidates in the operation set. 
$\alpha_k^{(i,j)}$ is the weight of the corresponding primitive operation $k$ on the edge $(i,j)$, and $p_{\alpha_k^{(i,j)}}$ is the probability of choosing this operation among $K$ candidates.
After neural architecture search, we can prune the redundant operations and paths by choosing $\alpha_{k}^{(i,j)}$ with the highest probability.

\subsection{Quality Control Network Search Space}\label{section_quality}

\begin{figure}[!htb]
	\centering
	\includegraphics[width=0.9\linewidth]{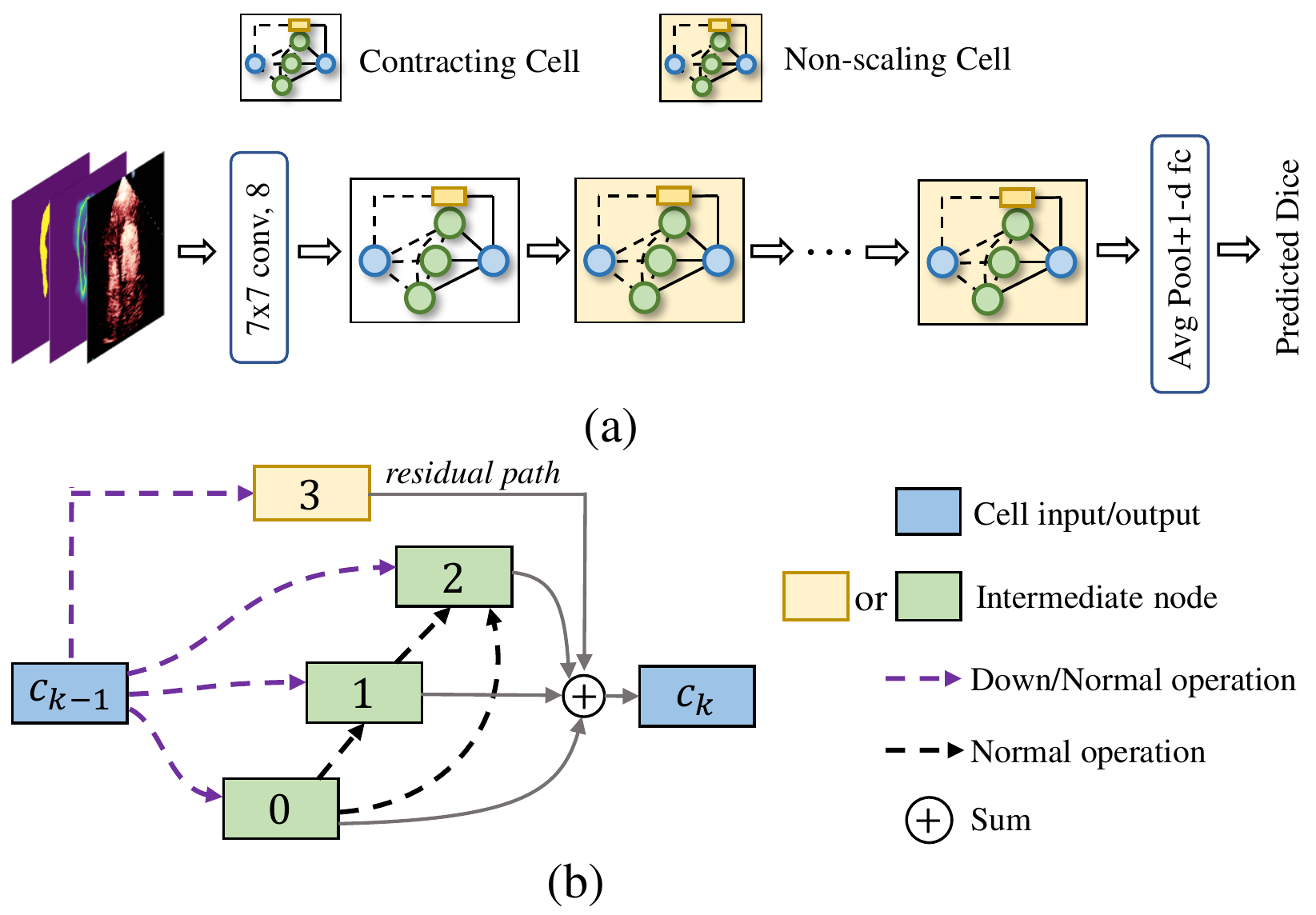}
	\caption{(a) The Resnet like quality control model. The network consists of two kinds of cell structures, namely Contrasting Cell and Non-scaling Cell. Two types of cells are stacked in an alternating way. (b) Detail cell structure. Dashed lines/edges represent the operations need to search. The purple edges denote the Down operation and Normal operation in the Contracting cell and Non-scaling cell respectively.}
	\label{fig:quality_control}
\end{figure}

Our quality control NAS network is shown in Fig.~\ref{fig:quality_control}(a).
In this network, we define two types of cell structures, namely Contracting cell and Non-scaling cell to minic the resnet backbone, they share the same network structure.
Fig.~\ref{fig:quality_control}(b) demonstrate the detail structure of the cells.
Unlike the cell structure in the segmentation network, the input of Contracting/Non-scaling cell is the cell output of the previous layer.
Inside a cell, a residual path (see Fig.~\ref{fig:quality_control}(b)) is used to mimic the residual behavior.
The down and normal operations, whose candidates are the same as in Section \ref{section_segmentation}, are used in Contracting cell and Non-scaling cell respectively .
The computation of feature map in the cell is also the same as in Section \ref{section_segmentation}.

\subsection{Hardware-aware Loss Function}\label{section_hardware}
In order to jointly optimize the segmentation accuracy, quality prediction accuracy and hardware latency, the following loss function is used during searching:
\begin{equation}
    min: \mathcal{L}(w,\alpha)=\mathcal{L}_{CE}(w,\alpha)+\lambda_1 \mathcal{L}_{MSE}(w,\alpha)+\lambda_2 LAT(\alpha).
\label{equ:total_loss}
\end{equation}
where $w$ is the network weights and $\alpha$ is the architecture parameters. 
The $\mathcal{L}_{CE}(w,\alpha)$ is the cross-entropy loss for the segmentation network. $\mathcal{L}_{MSE}(w,\alpha)$ is the mean squared error loss for the quality prediction network.
$LAT(\alpha)$ is the expected latency of the whole framework.
$\lambda_1$ and $\lambda_2$ are two scaling factors that control the trade-off between segmentation accuracy, quality prediction accuracy and hardware latency.
$\lambda_1$ and $\lambda_2$ are set to 1 and 0.001 in our experiments.

Similar to the latency estimation strategy in FBNet \cite{wu2019fbnet}, we use a differentiable method to estimate the expected latency of the cell and then the latency of the framework will be the summation of all the stacked cells in the network. 
Specifically, the latency of each candidate primitive operation is profiled into a lookup table for efficiency estimation.
During training, the estimated latency of each edge $(i,j)$ that needs to search in a cell is computed by summing up the weight latency of all candidate operations.
\begin{equation}
    latency(edge^{(i,j)})=\sum_{k\in K}GS(\alpha_{k}^{(i,j)}|\alpha^{(i,j)})latency(op_k^{(i,j)}).
\end{equation}
in which $GS(\cdot)$ is the Gumbel-Softmax sampling rule to make the architecture parameters differentiable to the latency loss term.
$latency(op_k^{(i,j)})$ is a constant coefficient from the lookup table.
Then, the estimated latency of a cell is the summation of the estimated latency of all edges, and the estimated latency of whole network is the summation of estimated latency of all cells.

\section{Experiments}

\subsection{Experimental Setup}

\textbf{Dataset and Evaluation Protocols.}
The dataset we use consists of 100 patients who received MCE from Guangdong Provincial People’s Hospital.
Each patient has an MCE sequence containing around 100 frames in A4C chamber view.
We randomly sample 30 frames from the MCE sequence of each patient, which results in 3000 images in total in this dataset.
All images are center cropped and resized to a fixed resolution 256$\times$256.
During searching, we split the 100 patients into $D_{train}$ and $D_{test}$ by a ratio of 8:2.
Since the quality control network needs to be trained on a dataset that is not used for training the segmentation network to avoid over-fitting, we further use 40 patients (denoted as $D_{seg\_train}$) in $D_{train}$ for training the segmentation network and use the rest 40 patients (denoted as $D_{seg\_eval}$) for evaluating the segmentation network.
The quality control network is trained on the whole $D_{train}$. When searching on $D_{seg\_eval}$, both networks are optimized using the loss in Equation~\ref{equ:total_loss}. When $D_{seg\_eval}$ is used, the segmentation network is frozen and only the parameters of the quality control network are updated.
Similar to \cite{liu2018darts}, we use the first-order approximation optimization strategy for searching, where the network weight parameters and architecture parameters are updated in turn.
After searching, for both networks, we randomly sample 60 subjects for training, the rest are used for validation.
The evaluation protocols include segmentation Dice, mean absolute error (MAE) of the predicted Dice, and the overall framework latency on CPU.
All the latency is emulated on a 4-Core Intel i7 processor which is used on the Philips Epiq 7 ultrasound machine.

\textbf{Setup.} In our NAS supernet, the intermediate nodes of all the cells are set to 3.
The initial filter size is set to 4 and 8 for segmentation network and quality control network for latency reduction.
Data augmentation such as translation, rotation and scale are used during searching. 
We search the network for 80 epochs in total, the first half is used to optimize the network weight.
For the rest 40 epochs, network weights and architecture parameters are optimized alternately.
We repeat the experiments five times with different random seeds and report the best validation performance of the architectures.
All experiments run on a machine with four NVIDIA GeForce GTX 1080 GPUs.

\subsection{Results}

\begin{table}[h]
\centering
\caption{Comparison between baseline results and our proposed hardware-aware NAS on Dice, MAE and latency. U-net(8) means the vanilla U-net with initial filter size 8.}
\resizebox{\columnwidth}{!}{
\begin{tabular}{lcccc}
\toprule
Method & Dice & MAE & Latency (ms) & FLOPs (M) \\\hline
U-net(8)$+$Resnet18 & 0.797 & 0.050 & 47 & 1045   \\
U-net(16)$+$Resnet18 & 0.839 & 0.049 & 82 & 3635  \\
w/o hardware-aware & 0.837 & 0.047 & 54 & 440 \\
hardware-aware & 0.832 & 0.049 & 49 & 373 \\ 
\bottomrule
\end{tabular}}
\label{table:result_baselines}
\end{table}

\begin{figure}[!htb]
	\centering
	\vspace{-7pt}
	\includegraphics[width=0.95\linewidth]{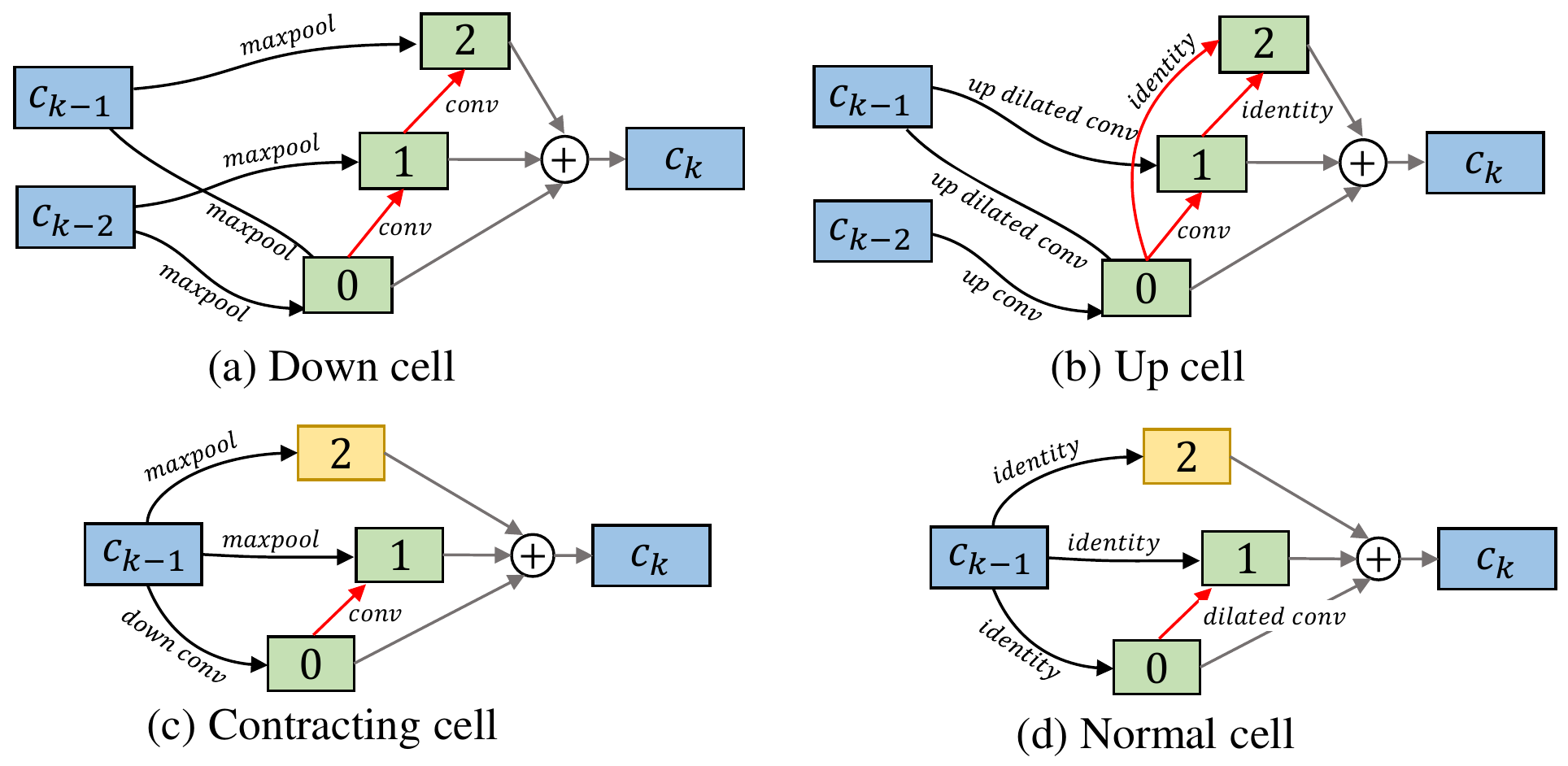}
	\caption{Illustration of searched cell structures, including the Down Cell, Up cell in segmentation network and the Contracting cell and Non-scaling cell in quality control network.}
	\vspace{-10pt}
	\label{fig:searched_cell}
\end{figure}

Table~\ref{table:result_baselines} shows the comparison results among the baselines and the proposed hardware-aware NAS.
It can be seen that compared to handcrafted neural architecture U-net(16)+Resnet18, the searched network reduces the inference latency by 40.2\% with very small accuracy degradation.
Integrating hardware latency during searching can further reduce the latency of the search architecture by 5ms.
All searched cell structures are shown in Fig. \ref{fig:searched_cell}.

\section{Conclusion}
In this paper, we proposed a hardware-aware NAS framework to jointly search for the optimal myocardial segmentation network and quality control network for contrast echocardiography with better accuracy and latency for the real-time requirement.
Experimental results on our MCE dataset show that searched networks can greatly reduce the inference time of the entire framework for real-time ultrasound acquisition guidance, while still maintaining high accuracy.

\bibliographystyle{IEEEtran}
\bibliography{IEEEabrv,IEEEexample}

\begin{thebibliography}{10}
\providecommand{\url}[1]{#1}
\csname url@samestyle\endcsname
\providecommand{\newblock}{\relax}
\providecommand{\bibinfo}[2]{#2}
\providecommand{\BIBentrySTDinterwordspacing}{\spaceskip=0pt\relax}
\providecommand{\BIBentryALTinterwordstretchfactor}{4}
\providecommand{\BIBentryALTinterwordspacing}{\spaceskip=\fontdimen2\font plus
\BIBentryALTinterwordstretchfactor\fontdimen3\font minus
  \fontdimen4\font\relax}
\providecommand{\BIBforeignlanguage}[2]{{%
\expandafter\ifx\csname l@#1\endcsname\relax
\typeout{** WARNING: IEEEtran.bst: No hyphenation pattern has been}%
\typeout{** loaded for the language `#1'. Using the pattern for}%
\typeout{** the default language instead.}%
\else
\language=\csname l@#1\endcsname
\fi
#2}}
\providecommand{\BIBdecl}{\relax}
\BIBdecl

\bibitem{li2017fully}
Y.~Li, C.~P. Ho, M.~Toulemonde, N.~Chahal, R.~Senior, and M.-X. Tang, ``Fully
  automatic myocardial segmentation of contrast echocardiography sequence using
  random forests guided by shape model,'' \emph{IEEE transactions on medical
  imaging}, vol.~37, no.~5, pp. 1081--1091, 2017.

\bibitem{li2021deep}
M.~Li, D.~Zeng \emph{et~al.}, ``A deep learning approach with temporal
  consistency for automatic myocardial segmentation of quantitative myocardial
  contrast echocardiography,'' \emph{The International Journal of
  Cardiovascular Imaging}, pp. 1--12, 2021.

\bibitem{dong2018voxelatlasgan}
S.~Dong \emph{et~al.}, ``Voxelatlasgan: 3d left ventricle segmentation on
  echocardiography with atlas guided generation and voxel-to-voxel
  discrimination,'' in \emph{International Conference on Medical Image
  Computing and Computer-Assisted Intervention}.\hskip 1em plus 0.5em minus
  0.4em\relax Springer, 2018, pp. 622--629.

\bibitem{roy2018inherent}
A.~G. Roy, S.~Conjeti, N.~Navab, and C.~Wachinger, ``Inherent brain
  segmentation quality control from fully convnet monte carlo sampling,'' in
  \emph{International Conference on Medical Image Computing and
  Computer-Assisted Intervention}.\hskip 1em plus 0.5em minus 0.4em\relax
  Springer, 2018, pp. 664--672.

\bibitem{robinson2019automated}
R.~Robinson, V.~V. Valindria, W.~Bai, O.~Oktay, B.~Kainz, H.~Suzuki, M.~M.
  Sanghvi, N.~Aung, J.~M. Paiva, F.~Zemrak \emph{et~al.}, ``Automated quality
  control in image segmentation: application to the uk biobank cardiovascular
  magnetic resonance imaging study,'' \emph{Journal of Cardiovascular Magnetic
  Resonance}, vol.~21, no.~1, pp. 1--14, 2019.

\bibitem{chen2017deeplab}
L.-C. Chen, G.~Papandreou \emph{et~al.}, ``Deeplab: Semantic image segmentation
  with deep convolutional nets, atrous convolution, and fully connected crfs,''
  \emph{IEEE transactions on pattern analysis and machine intelligence},
  vol.~40, no.~4, pp. 834--848, 2017.

\bibitem{ruijsink2020fully}
B.~Ruijsink \emph{et~al.}, ``Fully automated, quality-controlled cardiac
  analysis from cmr: validation and large-scale application to characterize
  cardiac function,'' \emph{Cardiovascular Imaging}, vol.~13, no.~3, pp.
  684--695, 2020.

\bibitem{hoebel2020exploration}
K.~Hoebel, V.~Andrearczyk, A.~Beers, J.~Patel, K.~Chang, A.~Depeursinge,
  H.~M{\"u}ller, and J.~Kalpathy-Cramer, ``An exploration of uncertainty
  information for segmentation quality assessment,'' in \emph{Medical Imaging
  2020: Image Processing}, vol. 11313.\hskip 1em plus 0.5em minus 0.4em\relax
  International Society for Optics and Photonics, 2020, p. 113131K.

\bibitem{jungo2018uncertainty}
A.~Jungo, R.~Meier, E.~Ermis, E.~Herrmann, and M.~Reyes, ``Uncertainty-driven
  sanity check: Application to postoperative brain tumor cavity segmentation,''
  \emph{arXiv preprint arXiv:1806.03106}, 2018.

\bibitem{annett2014low}
M.~Annett, A.~Ng, P.~Dietz, W.~F. Bischof, and A.~Gupta, ``How low should we
  go? understanding the perception of latency while inking,'' in
  \emph{Proceedings of Graphics Interface 2014}, 2014, pp. 167--174.

\bibitem{wang2019msu}
T.~Wang, J.~Xiong \emph{et~al.}, ``Msu-net: Multiscale statistical u-net for
  real-time 3d cardiac mri video segmentation,'' in \emph{International
  Conference on Medical Image Computing and Computer-Assisted
  Intervention}.\hskip 1em plus 0.5em minus 0.4em\relax Springer, 2019, pp.
  614--622.

\bibitem{robinson2018real}
R.~Robinson \emph{et~al.}, ``Real-time prediction of segmentation quality,'' in
  \emph{International Conference on Medical Image Computing and
  Computer-Assisted Intervention}.\hskip 1em plus 0.5em minus 0.4em\relax
  Springer, 2018, pp. 578--585.

\bibitem{jiang2019accuracy}
W.~Jiang \emph{et~al.}, ``Accuracy vs. efficiency: Achieving both through
  fpga-implementation aware neural architecture search,'' in \emph{Proceedings
  of the 56th Annual Design Automation Conference 2019}, 2019, pp. 1--6.

\bibitem{wu2019fbnet}
B.~Wu, X.~Dai, P.~Zhang \emph{et~al.}, ``Fbnet: Hardware-aware efficient
  convnet design via differentiable neural architecture search,'' in
  \emph{Proceedings of the IEEE/CVF Conference on Computer Vision and Pattern
  Recognition}, 2019, pp. 10\,734--10\,742.

\bibitem{cai2018proxylessnas}
H.~Cai \emph{et~al.}, ``Proxylessnas: Direct neural architecture search on
  target task and hardware,'' \emph{arXiv preprint arXiv:1812.00332}, 2018.

\bibitem{zhang2019fine}
R.~Zhang and A.~C. Chung, ``A fine-grain error map prediction and segmentation
  quality assessment framework for whole-heart segmentation,'' in
  \emph{International Conference on Medical Image Computing and
  Computer-Assisted Intervention}.\hskip 1em plus 0.5em minus 0.4em\relax
  Springer, 2019, pp. 550--558.

\bibitem{roy2019bayesian}
A.~G. Roy, S.~Conjeti \emph{et~al.}, ``Bayesian quicknat: model uncertainty in
  deep whole-brain segmentation for structure-wise quality control,''
  \emph{NeuroImage}, vol. 195, pp. 11--22, 2019.

\bibitem{valindria2017reverse}
V.~V. Valindria, I.~Lavdas, W.~Bai, K.~Kamnitsas, E.~O. Aboagye \emph{et~al.},
  ``Reverse classification accuracy: predicting segmentation performance in the
  absence of ground truth,'' \emph{IEEE transactions on medical imaging},
  vol.~36, no.~8, pp. 1597--1606, 2017.

\bibitem{elsken2019neural}
T.~Elsken, J.~H. Metzen, F.~Hutter \emph{et~al.}, ``Neural architecture search:
  A survey.'' \emph{J. Mach. Learn. Res.}, vol.~20, no.~55, pp. 1--21, 2019.

\bibitem{baker2016designing}
B.~Baker \emph{et~al.}, ``Designing neural network architectures using
  reinforcement learning,'' \emph{arXiv preprint arXiv:1611.02167}, 2016.

\bibitem{real2017large}
E.~Real, S.~Moore, A.~Selle, S.~Saxena, Y.~L. Suematsu \emph{et~al.},
  ``Large-scale evolution of image classifiers,'' in \emph{International
  Conference on Machine Learning}.\hskip 1em plus 0.5em minus 0.4em\relax PMLR,
  2017, pp. 2902--2911.

\bibitem{liu2018darts}
H.~Liu, K.~Simonyan, and Y.~Yang, ``Darts: Differentiable architecture
  search,'' \emph{arXiv preprint arXiv:1806.09055}, 2018.

\bibitem{weng2019unet}
Y.~Weng, T.~Zhou, Y.~Li, and X.~Qiu, ``Nas-unet: Neural architecture search for
  medical image segmentation,'' \emph{IEEE Access}, vol.~7, pp.
  44\,247--44\,257, 2019.

\end{thebibliography}


\end{document}